\documentclass[12pt]{elsarticle}
\usepackage[utf8]{inputenc}
\usepackage{amsmath, amssymb, latexsym}
\usepackage{float}
\usepackage{setspace}
\usepackage{tikz}
\usepackage{lineno,hyperref}

\makeatletter
    \def\ps@pprintTitle{%
       \let\@oddhead\@empty
       \let\@evenhead\@empty
       \def\@oddfoot{\reset@font\hfil\thepage\hfil}
       \let\@evenfoot\@oddfoot
    }
\makeatother

\modulolinenumbers[5]

\journal{Journal of Data\&Knowledge Engineering }
\usetikzlibrary{decorations.pathreplacing}
\usetikzlibrary{fadings}
\usetikzlibrary{positioning, fit, arrows.meta}
\usepackage{natbib}
\usepackage{graphicx}

\begin{document}

\begin{frontmatter}

\title{Leveraging Financial News for Stock Trend Prediction with Attention-Based Recurrent Neural Network}

\author{Huicheng Liu}
\address{Department of Electrical and Computer Engineering\\Queen's University, Canada\\Kingston, ON, Canada K7L 2N8}





\begin{abstract}
\normalsize
Stock market prediction is one of the most attractive research topic since the successful prediction on the market's future movement leads to significant profit. Traditional short term stock market predictions are usually based on the analysis of historical market data, such as stock prices, moving averages or daily returns. However, financial news also contains useful information on public companies and the market. 

Existing methods in finance literature exploit sentiment signal features, which are limited by not considering factors such as events and the news context. We address this issue by leveraging deep neural models to extract rich semantic features from news text. In particular, a Bidirectional-LSTM are used to encode the news text and capture the context information, self attention mechanism are applied to distribute attention on most relative words, news and days. In terms of predicting directional changes in both Standard \& Poor’s 500 index and individual companies stock price, we show that this technique is competitive with other state-of-the-art approaches, demonstrating the effectiveness of recent NLP technology advances for computational finance.

\end{abstract}

\end{frontmatter}

\section{Introduction}
Stock market prediction is the act of trying to determine the future value of a company stock~\cite{malkiel1985random}. Apparently, the successful prediction of a stock's future price can yield significant profit, making the prediction problem an area of strong appeal for both academic researchers and industry practitioners. However, Stock market prediction is usually considered as one of the most challenging issues among time series predictions due to its noise and volatile features~\cite{wang2012novel}. During the past decades, machine learning models, such as Support Vector Regression (SVR)~\cite{basak2007support} and Support Vector Machines(SVM)~\cite{hearst1998support}, have been widely used to predict financial time series and gain high predictive accuracy ~\cite{refenes1994stock,das2012support,lu2009financial}.

How to accurately predict stock movement is still an open question with respect to the economic and social organization of modern society. The well-known efficient-market hypothesis (EMH)~\cite{malkiel2003efficient} suggests that stock prices reflect all currently available information and any price changes based on the newly revealed relevant information. However, due to the implicit correlations between daily events and their effect on the stock prices, finding relevant information that contribute to the change of price for each individual stock are difficult. Besides, the influences of events to stock prices can occur in indirect ways and act in chain reactions, which sets obstacles for precise market prediction. There are three approaches related to the information required to make a prediction. The first approach, technical analysis, is based on the premise that the future behavior of a financial time series is conditioned to its own past. Secondly, fundamental analysis, is based on external information as political and economic factors. A major source of information are text from the internet, these information are taken from unstructured data as news articles, financial reports or even microblogs. Nofsinger et al~\cite{nofsinger2001impact} shows that in some cases, investors tend to buy after positive news resulting in a stress of buying and higher stock prices, they sell after negative news resulting in a decrease of prices. Finally the third approach considers as all relevant information coming from both, financial time series and textual data.

In this work, our goal is to leverage public released financial news and train a model named Attention-based LSTM (At-LSTM) to make prediction on directional changes for both Standard \& Poor’s 500 index and individual companies stock price. Our model consists a Recurrent Neural network(RNN) to encode the news text and capture the context information, self attention mechanism is applied to distribute attention on most relative words, news and days . The model input are financial news titles extracted from Reuters and Bloomberg. Our model take advantages from the rapid development of deep neural networks and we show that our model is competitive with other state-of-the-art approaches, demonstrating the effectiveness of recent NLP technology advances for computational finance.

The rest of the paper are organized as follows. We introduce related work in Section 2. Followed by, we present some background and the methodologies for our proposed prediction model in Section 3. Experimental setup and results are demonstrated in Section 4. The paper finally concludes and point out future directions in Section 5.
\section{Related work}
Stock market prediction is an intriguing time-series learning problem in finance and economics, which has attracted a considerable amount of research. Efforts on predicting stock market have been carried out based on different resources and approaches. For example, one of the most widely studied approach relies on analyzing recent prices and volumes on the market~\cite{schoneburg1990stock,akgiray1989conditional,goccken2016integrating,adebiyi2014comparison,kim2012simultaneous}. 

Analyzing stock market using relevant text is complicated but intriguing~\cite{wu2009stock,xie2013semantic,ding2014using,ding2015deep,ding2016knowledge,chang2016measuring,peng2015leverage,luss2015predicting,skuza2015sentiment,sehgal2007sops}. For instance, a model with the name Enalyst was introduced in Lavrenko et al. Their goal is to predict stock intraday price trends by analyzing news articles published in the homepage of YAHOO finance. Mittermayer and Knolmayer implemented several prototypes for predicting the short-term market reaction to news based on text mining techniques. Their model forecast 1-day trend of the five major companies indices. Wu et al.~\cite{wu2009stock} predicted stock trends by selecting a representative set of bursty features (keywords) that have impact on individual stocks. Vivek Sehgal et al.~\cite{sehgal2007sops} introduced a method to predict stock market using sentiment. Similarly, Michał Skuza et al.~\cite{skuza2015sentiment} used sentiment from postings on twitter to predict future stock prices. However, these methods have many limitations including unveiling the rules that may govern the dynamics of the market which makes the prediction model incapable to catch the impact of recent trends.  

More recently, neural networks have been leveraged to further improve the accuracy of prediction. In general, neural networks are able to learn dense representations of text, which have been shown to be effective on a wide range of NLP problems, given enough training sample. This is the case in stock market prediction where the prices of stocks are available together with a great collection of relevant text data. This provides a good setting for exploring deep learning-based models for stock price prediction. More specifically, dense representations can represent related sentiment, events, and factual information effectively, which can then be extremely challenging to represent using sparse indicator features.

Advance of deep learning models has inspired increasing efforts on stock market prediction by analyzing stock-related text. For example, Ding et al.~\cite{ding2015deep} showed that deep learning representation of event structures yields better accuracy compared to discrete event features. 
They further augmented their approach ~\cite{ding2016knowledge} to incorporating an outside knowledge graph into the learning process for event embedding. As another example, Chang et al.~\cite{chang2016measuring} used neural networks to directly learn representations for news abstracts, showing that it is effective for predicting the cumulative abnormal returns of public companies. Other work,e.g.,~\cite{kim2012simultaneous,goccken2016integrating,adebiyi2014comparison}, has proposed different models of neural network that improve the prediction accuracy. 

\section{Methodology}
In this section, we first introduce some relative background on our prediction models. Followed by, we introduce the design of our proposed model to predict the directional movements of Standard \& Poor’s 500 index and individual companies stock price using financial news titles. The model is named as Attention-based LSTM (At-LSTM) and shown in Fig~\ref{fig:traditional-convolutional-network}. 
\subsection{Background}
\subsubsection{Bag of Words Model and Word Embedding}
The Bag-of-Words model is a simplifying representation often used in Natural Language Processing and Information Retrieval~\cite{Bag}. Also known as the vector space model. In this model, a text (such as a sentence or a document) is represented as a bag of its words, disregarding grammar and even word order but keeping multiplicity. The bag-of-words model is commonly used in document classification where the occurrence of each word represents a feature for training a classifier.
 
Word Embedding is the collective name for a set of language modeling and feature learning techniques in Natural Language Processing (NLP) where words or phrases from the vocabulary are mapped to vectors of real numbers~\cite{Word}. Conceptually it involves a mathematical embedding from a space with one dimension per word to a continuous vector space with much lower dimension. Neural networks can be used to generate this mapping and the most common algorithm used are continuous bag of words(CBOW) and skip-gram algorithm~\cite{mikolov2013efficient}.

Word and phrase embeddings, when used as the underlying input representation, have been shown to boost the performance in NLP tasks such as syntactic parsing and sentiment analysis.

\subsubsection{Convolutional Neural Network}
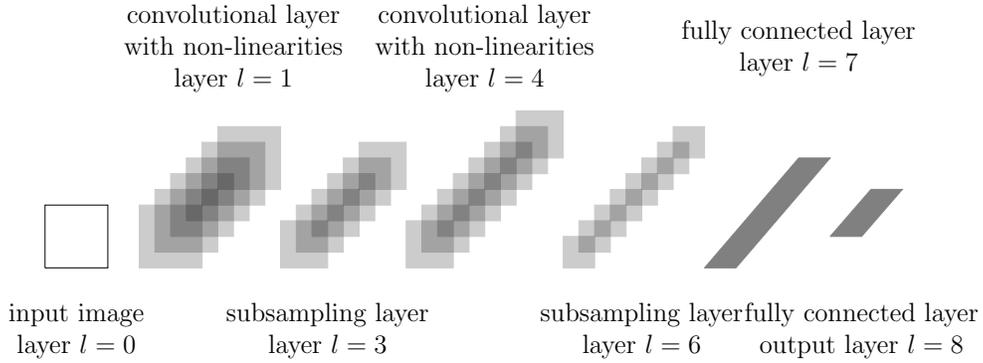
\begin{figure}[t!]
	\centering
	\resizebox{\columnwidth}{!}{
	\begin{tikzpicture}
		\node at (0.5,-1){\begin{tabular}{c}input image\\layer $l = 0$\end{tabular}};
		
		\draw (0,0) -- (1,0) -- (1,1) -- (0,1) -- (0,0);
		
		\node at (3,3.5){\begin{tabular}{c}convolutional layer\\with non-linearities\\layer $l = 1$\end{tabular}};
		
		\draw[fill=black,opacity=0.2,draw=black] (2.75,1.25) -- (3.75,1.25) -- (3.75,2.25) -- (2.75,2.25) -- (2.75,1.25);
		\draw[fill=black,opacity=0.2,draw=black] (2.5,1) -- (3.5,1) -- (3.5,2) -- (2.5,2) -- (2.5,1);
		\draw[fill=black,opacity=0.2,draw=black] (2.25,0.75) -- (3.25,0.75) -- (3.25,1.75) -- (2.25,1.75) -- (2.25,0.75);
		\draw[fill=black,opacity=0.2,draw=black] (2,0.5) -- (3,0.5) -- (3,1.5) -- (2,1.5) -- (2,0.5);
		\draw[fill=black,opacity=0.2,draw=black] (1.75,0.25) -- (2.75,0.25) -- (2.75,1.25) -- (1.75,1.25) -- (1.75,0.25);
		\draw[fill=black,opacity=0.2,draw=black] (1.5,0) -- (2.5,0) -- (2.5,1) -- (1.5,1) -- (1.5,0);
		
		\node at (4.5,-1){\begin{tabular}{c}subsampling layer\\layer $l = 3$\end{tabular}};
		
		\draw[fill=black,opacity=0.2,draw=black] (5,1.25) -- (5.75,1.25) -- (5.75,2) -- (5,2) -- (5,1.25);
		\draw[fill=black,opacity=0.2,draw=black] (4.75,1) -- (5.5,1) -- (5.5,1.75) -- (4.75,1.75) -- (4.75,1);
		\draw[fill=black,opacity=0.2,draw=black] (4.5,0.75) -- (5.25,0.75) -- (5.25,1.5) -- (4.5,1.5) -- (4.5,0.75);
		\draw[fill=black,opacity=0.2,draw=black] (4.25,0.5) -- (5,0.5) -- (5,1.25) -- (4.25,1.25) -- (4.25,0.5);
		\draw[fill=black,opacity=0.2,draw=black] (4,0.25) -- (4.75,0.25) -- (4.75,1) -- (4,1) -- (4,0.25);
		\draw[fill=black,opacity=0.2,draw=black] (3.75,0) -- (4.5,0) -- (4.5,0.75) -- (3.75,0.75) -- (3.75,0);
		
		\node at (7,3.5){\begin{tabular}{c}convolutional layer\\with non-linearities\\layer $l = 4$\end{tabular}};
		
		\draw[fill=black,opacity=0.2,draw=black] (7.5,1.75) -- (8.25,1.75) -- (8.25,2.5) -- (7.5,2.5) -- (7.5,1.75);
		\draw[fill=black,opacity=0.2,draw=black] (7.25,1.5) -- (8,1.5) -- (8,2.25) -- (7.25,2.25) -- (7.25,1.5);
		\draw[fill=black,opacity=0.2,draw=black] (7,1.25) -- (7.75,1.25) -- (7.75,2) -- (7,2) -- (7,1.25);
		\draw[fill=black,opacity=0.2,draw=black] (6.75,1) -- (7.5,1) -- (7.5,1.75) -- (6.75,1.75) -- (6.75,1);
		\draw[fill=black,opacity=0.2,draw=black] (6.5,0.75) -- (7.25,0.75) -- (7.25,1.5) -- (6.5,1.5) -- (6.5,0.75);
		\draw[fill=black,opacity=0.2,draw=black] (6.25,0.5) -- (7,0.5) -- (7,1.25) -- (6.25,1.25) -- (6.25,0.5);
		\draw[fill=black,opacity=0.2,draw=black] (6,0.25) -- (6.75,0.25) -- (6.75,1) -- (6,1) -- (6,0.25);
		\draw[fill=black,opacity=0.2,draw=black] (5.75,0) -- (6.5,0) -- (6.5,0.75) -- (5.75,0.75) -- (5.75,0);
		
		\node at (9.5,-1){\begin{tabular}{c}subsampling layer\\layer $l = 6$\end{tabular}};
		
		\draw[fill=black,opacity=0.2,draw=black] (10,1.75) -- (10.5,1.75) -- (10.5,2.25) -- (10,2.25) -- (10,1.75);
		\draw[fill=black,opacity=0.2,draw=black] (9.75,1.5) -- (10.25,1.5) -- (10.25,2) -- (9.75,2) -- (9.75,1.5);
		\draw[fill=black,opacity=0.2,draw=black] (9.5,1.25) -- (10,1.25) -- (10,1.75) -- (9.5,1.75) -- (9.5,1.25);
		\draw[fill=black,opacity=0.2,draw=black] (9.25,1) -- (9.75,1) -- (9.75,1.5) -- (9.25,1.5) -- (9.25,1);
		\draw[fill=black,opacity=0.2,draw=black] (9,0.75) -- (9.5,0.75) -- (9.5,1.25) -- (9,1.25) -- (9,0.75);
		\draw[fill=black,opacity=0.2,draw=black] (8.75,0.5) -- (9.25,0.5) -- (9.25,1) -- (8.75,1) -- (8.75,0.5);
		\draw[fill=black,opacity=0.2,draw=black] (8.5,0.25) -- (9,0.25) -- (9,0.75) -- (8.5,0.75) -- (8.5,0.25);
		\draw[fill=black,opacity=0.2,draw=black] (8.25,0) -- (8.75,0) -- (8.75,0.5) -- (8.25,0.5) -- (8.25,0);
		
		\node at (12,3.5){\begin{tabular}{c}fully connected layer\\layer $l = 7$\end{tabular}};
		
		\draw[fill=black,draw=black,opacity=0.5] (10.5,0) -- (11,0) -- (12.5,1.75) -- (12,1.75) -- (10.5,0);
		
		\node at (13,-1){\begin{tabular}{c}fully connected layer\\output layer $l = 8$\end{tabular}};
		
		\draw[fill=black,draw=black,opacity=0.5] (12.5,0.5) -- (13,0.5) -- (13.65,1.25) -- (13.15,1.25) -- (12.5,0.5);
	\end{tikzpicture}
	}
	\caption[Architecture of a traditional convolutional neural network.]{The architecture of the original convolutional neural network, as introduced by LeCun et al. (1989), alternates between convolutional layers including hyperbolic tangent non-linearities and subsampling layers. In this illustration, the convolutional layers already include non-linearities and, thus, a convolutional layer actually represents two layers. The feature maps of the final subsampling layer are then fed into the actual classifier consisting of an arbitrary number of fully connected layers. The output layer usually uses softmax activation functions.}
	\label{fig:traditional-convolutional-network}
\end{figure}

As shown in figure~\ref{fig:traditional-convolutional-network}, Convolutional Neural Network (CNN) is a class of deep, feed-forward artificial neural networks that has successfully been applied to analyzing visual imagery and NLP related tasks. Same as neural network, CNN are made up of neurons that have learnable weights and biases. Each neuron receives some inputs, performs a dot product and optionally follows it with a non-linearity. A Convolutional layer is composed by four consecutive operations: Convolution, subsampling(pooling), activation and dropout. The convolution layer can help select features so that CNN requires minimal pre-processing compared to other deep learning models. The core part of the Convolutional neural network is the convolution filter.

Convolution filters leverage two important ideas that can help improve a machine learning system: sparse interaction and parameter sharing. Sparse interaction contrasts with traditional neural networks where each output is interactive with each input. As a result, the output is only interactive with a narrow window of the input. Parameter sharing refers to reusing the filter parameters in the convolution operations, while the element in the weight matrix of traditional neural networks are used only once to calculate the output.

\subsubsection{Long-Short Term Memory}
Long Short Term Memory (LSTM) networks are a special kind of RNN, proposed by Hochreiter in 1997~\cite{hochreiter1997long}. LSTM units are a building unit for layers of a Recurrent Neural Network(RNN)~\cite{LSTM}. A RNN composed of LSTM units are often called an LSTM network. LSTM are explicitly designed to avoid the long-term dependency problem. A common LSTM unit is composed of a cell, an input gate, an output gate and a forget gate. The cell is responsible for "remembering" values over arbitrary time intervals; hence LSTM is capable of learning long-term dependencies.
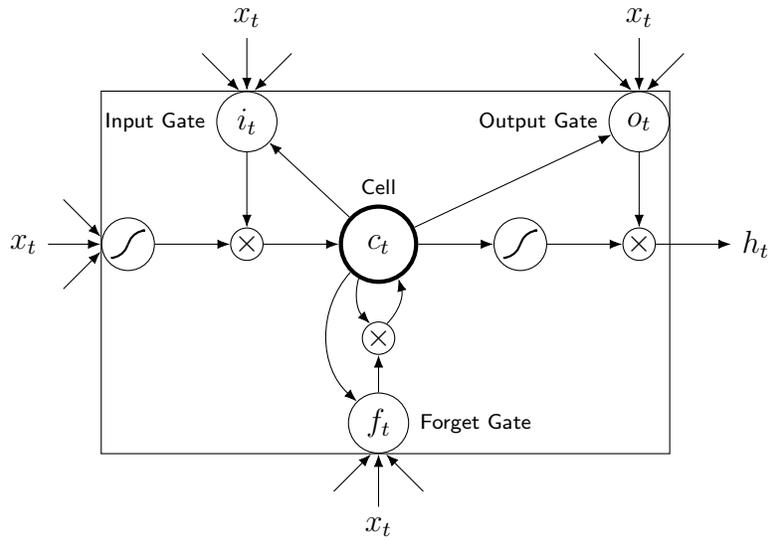
\begin{figure}[H]
    \centering

\begin{tikzpicture}[
    prod/.style={circle, draw, inner sep=0pt},
    ct/.style={circle, draw, inner sep=5pt, ultra thick, minimum width=10mm},
    ft/.style={circle, draw, minimum width=8mm, inner sep=1pt},
    filter/.style={circle, draw, minimum width=7mm, inner sep=1pt, path picture={\draw[thick, rounded corners] (path picture bounding box.center)--++(65:2mm)--++(0:1mm);
    \draw[thick, rounded corners] (path picture bounding box.center)--++(245:2mm)--++(180:1mm);}},
    mylabel/.style={font=\scriptsize\sffamily},
    >=LaTeX
    ]

\node[ct, label={[mylabel]Cell}] (ct) {$c_t$};
\node[filter, right=of ct] (int1) {};
\node[prod, right=of int1] (x1) {$\times$}; 
\node[right=of x1] (ht) {$h_t$};
\node[prod, left=of ct] (x2) {$\times$}; 
\node[filter, left=of x2] (int2) {};
\node[prod, below=5mm of ct] (x3) {$\times$}; 
\node[ft, below=5mm of x3, label={[mylabel]right:Forget Gate}] (ft) {$f_t$};
\node[ft, above=of x2, label={[mylabel]left:Input Gate}] (it) {$i_t$};
\node[ft, above=of x1, label={[mylabel]left:Output Gate}] (ot) {$o_t$};

\foreach \i/\j in {int2/x2, x2/ct, ct/int1, int1/x1,
            x1/ht, it/x2, ct/it, ct/ot, ot/x1, ft/x3}
    \draw[->] (\i)--(\j);

\draw[->] (ct) to[bend right=45] (ft);

\draw[->] (ct) to[bend right=30] (x3);
\draw[->] (x3) to[bend right=30] (ct);

\node[fit=(int2) (it) (ot) (ft), draw, inner sep=0pt] (fit) {};

\draw[<-] (fit.west|-int2) coordinate (aux)--++(180:7mm) node[left]{$x_t$};
\draw[<-] ([yshift=1mm]aux)--++(135:7mm);
\draw[<-] ([yshift=-1mm]aux)--++(-135:7mm);

\draw[<-] (fit.north-|it) coordinate (aux)--++(90:7mm) node[above]{$x_t$};
\draw[<-] ([xshift=1mm]aux)--++(45:7mm);
\draw[<-] ([xshift=-1mm]aux)--++(135:7mm);

\draw[<-] (fit.north-|ot) coordinate (aux)--++(90:7mm) node[above]{$x_t$};
\draw[<-] ([xshift=1mm]aux)--++(45:7mm);
\draw[<-] ([xshift=-1mm]aux)--++(135:7mm);

\draw[<-] (fit.south-|ft) coordinate (aux)--++(-90:7mm) node[below]{$x_t$};
\draw[<-] ([xshift=1mm]aux)--++(-45:7mm);
\draw[<-] ([xshift=-1mm]aux)--++(-135:7mm);
\end{tikzpicture}
    \caption[Architecture of a traditional Long short term memory neural network.]{The architecture of the original Long short term memory neural network, as introduced by Hochreiter et al. ~\cite{hochreiter1997long},  Here, $f_t$ is the forget gate's activation vector, $i_t$ is the input gate's activation vector, $o_t$ is the output gate's activation vector, $x_t$ is the input vector to LSTM unit, $h_t$ is the output vector of the LSTM unit and $C_t$ is the cell state vector}
    \label{fig:traditional LSTM neural network}
\end{figure}
The expression long-short term refers to the fact that LSTM is a model for the short-term memory which can last for a long period of time. An LSTM is well-suited to classify, process and predict time series given time lags of unknown size and duration between important events. Besides, LSTM shows the promising result in sentence encoding in many NLP applications ~\cite{sutskever2014sequence,gers2002learning}. The LSTM archticture are shown in figure~\ref{fig:traditional LSTM neural network}, the computations of LSTM cells are:
\begin{spacing}{1.0}
\begin{align}
	f_t = \sigma(W_f[h_{t-1},x_t] + b_f)
\end{align}
\begin{align}
	 i_t = \sigma(W_i[h_{t-1},x_t] + b_i)
\end{align}
\begin{align}
	\tilde{C}_t = tanh(W_C[h_{t-1}, x_t] + b_C)
\end{align}
\begin{align}
	C_t = f_t \otimes C_{t-1} + i_t \otimes \tilde{C}_t
\end{align}
\begin{align}
	 o_t = \sigma(W_o[h_{t-1},x_t] + b_o)
\end{align}
\begin{align}
	h_t = o_t \otimes tanh(C_t)
\end{align}
\end{spacing}

The forget gate is described in Equation (1) and is used to decide whether a value should be reserved. The Input gate in Equation (2) is used to control which values should be updated through a certain time step. Here, a new value $\tilde{C}_t$ is created as per Equation (3) using the $tanh$ function. Next, with Equation (4), the cell state will be updated from $C_{t-1}$ to $C_t$, $f_t$ and $i_t$ are used here to decide whether an information should be discarded. in Equation (5), an output gate is used to filter out redundant values. The final output is calculated by Equation (6). $W$ is the weight matrix and $b$ refers to the bias.

\subsubsection{Optimization Algorithm and Loss Function}
In this section, we will briefly introduce the Optimization algorithm and loss function we used in our prediction model. 

\begin{itemize}
  \item \textit{\textbf{Adadelta Optimization Algorithm}}: Apart from Stochastic Gradient Descent, Adaptive Gradient Algorithm or the famous Adam algorithm~\cite{kingma2014adam}, we chose Adadelta~\cite{zeiler2012adadelta} as our optimization algorithm. Adadelta is an optimization algorithm that can be used to update network weights iterative based on training data. Adadelta combines the advantages of two other extensions of stochastic gradient descent, Adaptive Gradient Algorithm and Root Mean Square Propagation Algorithm. The method dynamically adapts over time using only first order information and has minimal computational overhead beyond vanilla Stochastic Gradient Descent~\cite{zeiler2012adadelta}. The method requires no manual tuning of a learning rate and appears robust to noisy gradient information, different model architecture choices, various data modalities and selection of hyperparameters.
 \item \textit{\textbf{Cross Entropy Loss}}: In information theory, the cross entropy between two probability distributions $p$ and $q$ over the same underlying set of events measures the average number of bits needed to identify an event drawn from the set~\cite{Cross}. Cross entropy is the most suitable loss function since we want to measure the output from our prediction model with it's true output. A binary cross entropy loss formula are shown in Equation (7). In which, $J(w)$ refers to the loss, $N$ refers to the number of examples, $y_n$ is the expected output and $\tilde{y}_n$ is true output. We use this formula as our loss function in the model.
\begin{align}
    J(w)=-\frac{1}{N}H(p_n,q_n)=-\frac{1}{N}\displaystyle\sum_{i=1}^{N} [y_n*\log\tilde{y}_n+(1-y_n)*\log(1-\tilde{y}_n)]
\end{align}
\end{itemize}
\subsection{Model design}
\begin{figure}[H]
\centering
\includegraphics[width=1.0\textwidth,height=3.5in]{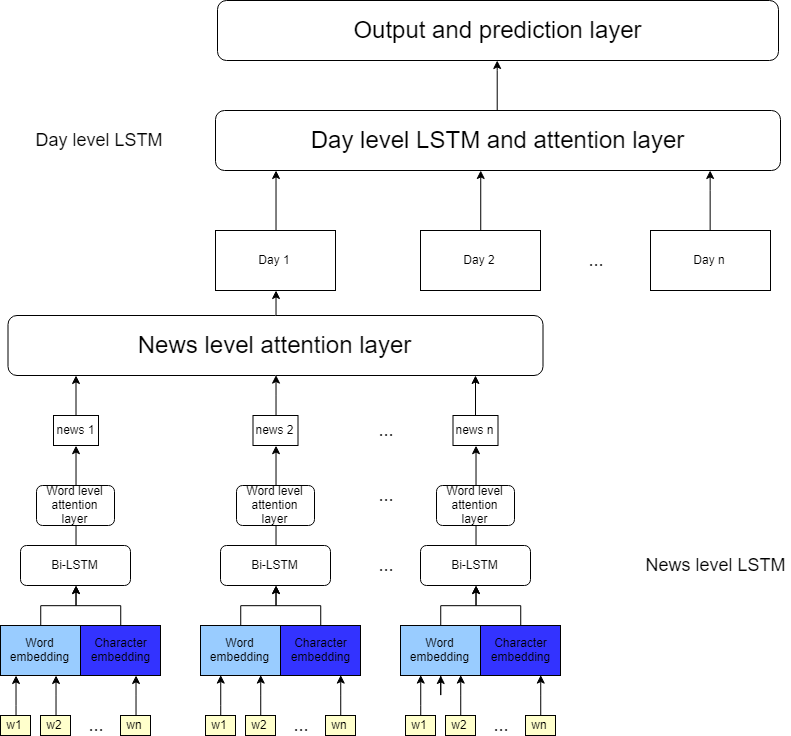}
\caption{\label{fig:model}Model Structure}
\end{figure}

In this subsection, we introduce the design of our Attention-based LSTM model (At-LSTM). The design has four stages: Input and Embedding layer, news-level Bi-LSTM and self-attention layer, day-level Bi-LSTM and self-attention layer and the final output and prediction layer. These stages will be described below and shown in figure ~\ref{fig:model}.

\subsubsection{Input and Embedding Layer}
\textit{\textbf{Model Input}}: The same world event can be described in different expression ways. This variety makes the representation space for events very sparse. Several approaches~\cite{ding2015deep,ding2016knowledge} represent the event as tuple $< S; V; O >$ to gain generalization. However, we argue that this representation is oversimplified and might lose lots of valuable information. Instead, we use the entire news title content as our models input and use LSTM-based encoder to encode it to a distributional representation to tackle with sparsity problem.
\\\textit{\textbf{Word and Character Embedding}}: For each input news, we remove the punctuation and then use a word embedding layer to calculate the embedded vectors for each word. The embedding layer takes a sequence of sentences as input, this sequence corresponds to a set of titles of news articles.
These embedding are unique vectors of continuous values with length $w = (w_1,...,w_l)$ and \begin{math}w_i\in R^{m}\end{math} for each word in the training corpus, $m$ is the word level embedding dimension. 

Existing pre-trained word embedding such as Glove~\cite{pennington2014glove} and Word2Vec~\cite{mikolov2013efficient} typically comes from general domains( Google News or Wikipedia, etc). However, these word embeddings often fail to capture rich domain specific vocabularies. We therefore train our own word embedding with financial domain news text consisting of news articles from Reuters an Bloomberg. The data are further described in Section 4.

In addition, we leverage character composition from Chen et al.~\cite{chen2017recurrent} and concatenate the character level composition with the original word embedding to gain rich represent for each word. The character composition feeds all characters of each word into a Convolutional Neural Network (CNN) with max-pooling~\cite{kim2014convolutional} to obtain representations $c = (c_1,...,c_l)$ and \begin{math}c_n\in R^{n}\end{math} for each word in the training corpus, $n$ is the character composition dimension. Finally, each word is represented as a concatenation of word-level embedding and character-composition vector $e_i = [w_i;c_i]$. A matrices \begin{math}e^s\in R^{k*(m+n)}\end{math} can be used to represent a news after the embedding layer, where $k$ is the length of the news.

\subsubsection{News Level Bi-LSTM and Self-Attention Layer}
\textit{\textbf{Bi-LSTM Encoding}}: After the embedding layer, we fed the words and their context in the news title into a Bi-LSTM based sentence encoder to perform distributional representation.  Bidirectional LSTM (Bi-LSTM) is a variant of LSTM which shows better result then uni-direction LSTM in recent NLP tasks as they can understand context better. A bidirectional LSTM runs a forward and backward LSTM on a sequence starting from the left and the right end, respectively~\cite{chen2017recurrent}. In this case, Bi-LSTM can not only preserve information from the past, but also catch the information from the future. We obtain the hidden vectors($\overrightarrow{h_i}$ and $\overleftarrow{h_i}$ shown in Equation (8) and (9)) from the sentence encoders and concatenate them to $H^t=[\overrightarrow{h_1};\overleftarrow{h_1},....,\overrightarrow{h_m};\overleftarrow{h_m}]$, \begin{math}H_i^t\in R^{2u}\end{math} represents the $ith$ news title after encoding in date $t$ and $m$ refers to the sequence number.

\begin{spacing}{1.0}
\begin{align}
	\overrightarrow{h_n} = \overrightarrow{LSTM}(e^s)
\end{align}
\begin{align}
	\overleftarrow{h_n} = \overleftarrow{LSTM}(e^s)
\end{align}
\end{spacing}

\textit{\textbf{Word level Self-attention layer}}: Instead of taking the average of the hidden vector after the sentence encoding, we leverage multi-hop self-attention mechanism~\cite{lin2017structured} on top of the Bi-LSTM layer. The attention mechanism takes the whole LSTM hidden states $H^t_i$ as input, and outputs a vector of weights $A$:

\begin{align}
    A = softmax(W_2tanh(W_1H_i^{t\intercal}))
\end{align}

Shown in figure~\ref{fig:attention}, here $W_1$ is a weight matrix with a shape of \begin{math}W_1\in R^{d_a*2u}\end{math} which u refers to the hidden unit of the news level Bi-LSTM. and $W_2$ is a vector of parameters with size \begin{math}W_2\in R^{r*d_a}\end{math}, $d_a$ and $r$ are hyper parameter that can be set arbitrarily. $H_i^t$ are sized \begin{math}H_i^t\in R^{n*2u}\end{math}, and the annotation vector $A$ will have a size \begin{math}A\in R^{r*n}\end{math}, the $softmax()$ ensures all the computed weights sum up to 1. Then we sum up the LSTM hidden states $H_i^t$ according to the weight provided by $A$ to get a vector representation $N_i^t$ for the input sentence. We can deem Equation (10) as a 2-layer MLP without bias, whose hidden units numbers are $d_a$ and parameters are ${W_1,W_2}$. We compute the $r$ weighted sums by multiplying the annotation matrix $A$ and LSTM hidden states $H_i^t$:

\begin{figure}[H]
\centering
\includegraphics[width=1.0\textwidth,height=4.0in]{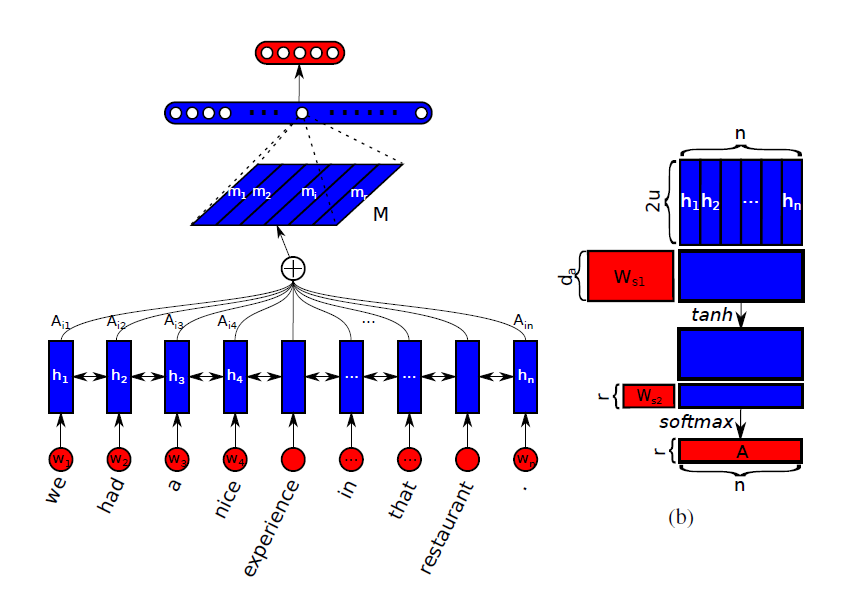}
\caption{\label{fig:attention}Self attention mechanism}
\end{figure}

\begin{align}
    N_i^t = AH_i^t
\end{align}

Eventually, the sentence encoding vector $H_i^t$ then becomes a matrix \begin{math}N_i^t\in R^{r*2u}\end{math} and we use $N_i^t$ to represent the $ith$ news title in date $t$ after encoding. By using multi-hop attention mechanism, the vector representation usually focuses on a specific component of the sentence, like a special set of related words or phrases. Therefore, it is expected to reflect an aspect, or component of the semantics in a sentence instead of adding attention on a specific word.

Shown in equation (12), we apply another MLP layer on top of our self-attention layer to learn which attention group should be rewarded with the highest assign value. We name this as attention-over-attention, the weight matrix have a shape of \begin{math}W_3\in R^r\end{math}, and the final representation of the sentence encoding vector $N_i^t$ are shaped as \begin{math}N_i^t\in R^{2u}\end{math}. At last, we use $N_i^t$to represent a news title encoded from the input.

\begin{align}
    N_i^t = tanh(W_3N_i^t+b_1)
\end{align}

\textit{\textbf{News level Self-attention layer}}:
Not all news contributes equally to predicting the stock trend. Hence, in order to reward the news that offers critical information, we apply the same structure multi-hop self-attention on top of the encoding layer to aggregate the news weighted by an assigned attention value. Specifically:
\begin{spacing}{1.0}
\begin{align}
    A = softmax(W_5tanh(W_4N^{t\intercal}))
\end{align}
\begin{align}
   D_t = AN^t
\end{align}
\begin{align}
    D_t = tanh(W_6D_t+b_2)
\end{align}
\end{spacing}

Here, $N^t=(N_1^t,....,N_m^t)$ and \begin{math}N^t\in R^{m*2u}\end{math}, in which $m$ refers to the number of news in date $t$. Note that the weights $[{W_4,W_5,W_6,b_2}]$ in the news level attention layer are different from the word level attention layer weight $[{W_1,W_2,W_3,b_1}]$. A vector $D_t$ represents the temporal sequence for all the news proposed in the date $t$. The merit of using multi-hop self-attention mechanism is that it learns and assign different groups of attention value to the news encoding. Formally, the first group of attention reward the news that contains positive sentiments to the stock market("raise", "growth" or "decrease", "down" etc). Whereas the second group of attention assign there reward to the news that mentions the major companies in the S\&P 500 ("Microsoft","Google" instead of a small company outside of the S\&P 500).

Obviously, the attention layer can be trained end-to-end and thus gradually learn to assign more attention to the reliable and informative news based on its content.
\subsubsection{Day level Bi-LSTM and self-attention layer}
\begin{spacing}{1.0}
\begin{align}
	\overrightarrow{h_i} = \overrightarrow{LSTM}(D_t)
\end{align}
\begin{align}
	\overleftarrow{h_i} = \overleftarrow{LSTM}(D_t)
\end{align}
\end{spacing}
We adopt day level Bi-LSTM to encode the temporal sequence of corpus vectors $D_i$, \begin{math}t\in [1,N]\end{math}. Shown in Equation (16) and (17), We obtain the hidden vectors($\overrightarrow{h_i}$ and $\overleftarrow{h_i}$) from the day-level Bi-LSTM and concatenate them to $H_i=[\overrightarrow{h_1};\overleftarrow{h_1},....,\overrightarrow{h_N};\overleftarrow{h_N}]$, $H_i$ represents a vector that encodes the temporal sequence $D_t$ where $N$ refers to the sequence number.
Since the news published at different dates contribute to the stock trend unequally, we adopt self-attention mechanism again to reward the dates that contribute most to the stock trend prediction, Shown in Equation below:
\begin{spacing}{1.0}
\begin{align}
    A = softmax(W_8tanh(W_7H_i^\intercal))
\end{align}
\begin{align}
    V = AD
\end{align}
\begin{align}
    V = tanh(W_9V+b_3)
\end{align}
\end{spacing}

In the formula, $D=(D_1,....,D_t)$ and \begin{math}D\in R^{N*2v}\end{math} , \begin{math}V\in R^{2v}\end{math} represents the final vector for all news proposed before the prediction date $t+1$ in a delay window with size $N$, where $v$ is the hidden unit number in the day level Bi-LSTM. Note that the weight matrix $[{W_7,W_8,W_9,b_3}]$ in the day level attention layer are different from the weight matrices mentioned in the previous section. 

\subsubsection{Output and Prediction Layer}
The last stage of our At-LSTM model is a traditional fully connected layer with softmax as activation function whose output is the probability distribution over labels. In this work, the objective is to forecast the direction of daily price movements of the stock price, this direction are used to create a binary class label where a label [1,0] represents that the stock price will increase and label [0,1] represents that the stock price will decrease.

\section{Experiments}
\subsection{Experimental Setup}
\subsubsection{Data}

We evaluated our model on a data set of financial news collected from Reuters and Bloomberg over the time period from October 2006 to November 2013. This data set was made publicly available by Ding et al.~\cite{ding2014using} and shown in table~\ref{tab:Data set}. We further collected data from Reuters for 473 companies listed in the Standard \& Poor's 500 over the time period started from November 2013 to march 2018. Meanwhile, the historical stock price data from October 2006 to March 2018 for all individual shares in Standard \& Poor's 500 are collected from Yahoo Finance. The second part of the data are used for individual stock price prediction and shown in table ~\ref{tab:individual}. Due to the page limit, We only show the major companies listed in the S\&P 500, this will also be applied in the result section.

\begin{table}[H]
\centering
\begin{tabular}{ |p{3cm}||p{3cm}|p{3cm}|p{3cm}| }
 \hline
 \multicolumn{4}{|c|}{Data for S\&P 500 index prediction} \\
 \hline
 Data set & Training & Development & Testing\\
 \hline
Time interval & 20/10/2006-27/06/2012 & 28/06/2012-13/03/2013&   14/03/2013-20/11/2013\\
\hline
News&  445,262 & 55,658 & 55,658\\
\hline
\end{tabular}\caption{\label{tab:Data set}Data for S\&P 500 index prediction}
\end{table}

\begin{table}
\centering
\begin{tabular}{ |p{3cm}||p{3cm}|p{3cm}|p{3cm}| }
 \hline
 \multicolumn{4}{|c|}{Data for individual stock prediction} \\
 \hline
 Company Symbol & Training news& Development news& Testing news\\
 \hline
GOOG & 5744 & 718 & 718\\
\hline
AMZN & 3245 & 406 & 405\\
\hline
CSCO & 2471 & 309 & 308\\
 \hline
MSFT & 4056 & 406 & 405\\
\hline
AAPL & 9720 & 1215 & 1214\\
 \hline
INTC & 4355 & 545 & 544\\
\hline
IBM & 2016 & 252 & 252\\
 \hline
AMD & 1224 & 153 & 153\\
\hline
NVDA & 1440 & 168 & 167\\
\hline
QCOM & 924 & 116 & 115\\
\hline
WMT& 2793 & 350 & 349\\
\hline
T& 1228 & 154 & 153\\
\hline
\end{tabular}\caption{\label{tab:individual}Data for individual stock prediction: The time interval of the news crawled from Reuters various between different companies. Hence, here we only present the number of articles in the training, development and testing set respectively.}
\end{table}

Following Ding et al.~\cite{ding2014using} we focus on the news headlines instead of the full content of the news articles for prediction since they found it produced better results. We use news articles on each day and predict if the S\&P 500 index closing price movement (increase or decrease) in the day $t+1$ compared with the closing price on day $t$.


\subsubsection{Implementation Details}
As mentioned in the previous section, We pre-trained 100 dimentional word embedding~\cite{mikolov2013efficient} with skip-gram algorithm on the data set shown in table~\ref{tab:Data set}, the size of the trained vocabulary is 153,214. In addition, firm names and an UNK token to represent any words out of the vocabulary are added to the vocabulary set, having an initial embedding initialized randomly with Gaussian samples. The word embedding are fine-tuned during model training. The character embedding has 15 dimensions, and CNN filters length are [1,3,5] respectively, each of those are 32 dimensions. The news level Bi-LSTM and day-level Bi-LSTM both have 300 hidden units, the day level LSTM window size $N$ has been set to 7. Hyper-parameters $d_a$ and $r$ in the self attention layer are set to 600 and 10, respectively. Mentioned in section 3, we use Adadelta for our optimization algorithm, the initial learning rate has been set to 0.04. Our model are trained for 200 Epoch.

\subsection{Base Lines and Proposed Model}
In this subsection, we propose a few baselines to compare with our proposed model. For the sake of simplicity, the following notation identifies each model:
\begin{itemize}
    \item \textit{\textbf{SVM}}: Luss and d’Aspremont et al. ~\cite{luss2015predicting} propose using Bags-of-Words to represent news documents, and constructed the prediction model using Support Vector Machines (SVMs).
    \item \textit{\textbf{Bag-At-LSTM}}: At-LSTM without sentence encoder. We take the average of the word embedding inputs instead of using Bi-LSTM to encode the news title.
    \item \textit{\textbf{WEB-At-LSTM}}: Same as our proposed model but without the character level composition.
    \item \textit{\textbf{Ab-At-LSTM}}: Instead of the news title, we use the news abstract as input for our model, the model structure remains the same.
    \item \textit{\textbf{Doc-At-LSTM}}: We further leverage a Hierarchical Attention Networks proposed by Yang et al.~\cite{yang2016hierarchical} to represent the entire news Document, this model adds a sentence level attention layer into our proposed model to differentiate more and less important content when constructing the document representation. 
    \item \textit{\textbf{Tech-At-LSTM}}: We concatenate seven Technical indicator leveraged from Zhai et al.~\cite{zhai2007combining} shown in figure ~\ref{fig:tech} with the vector $V$ after the day level LSTM layer and fed together into the prediction layer.
    
    \begin{figure}[H]
    \centering
    \includegraphics[width=1.0\textwidth,height=3.0in]{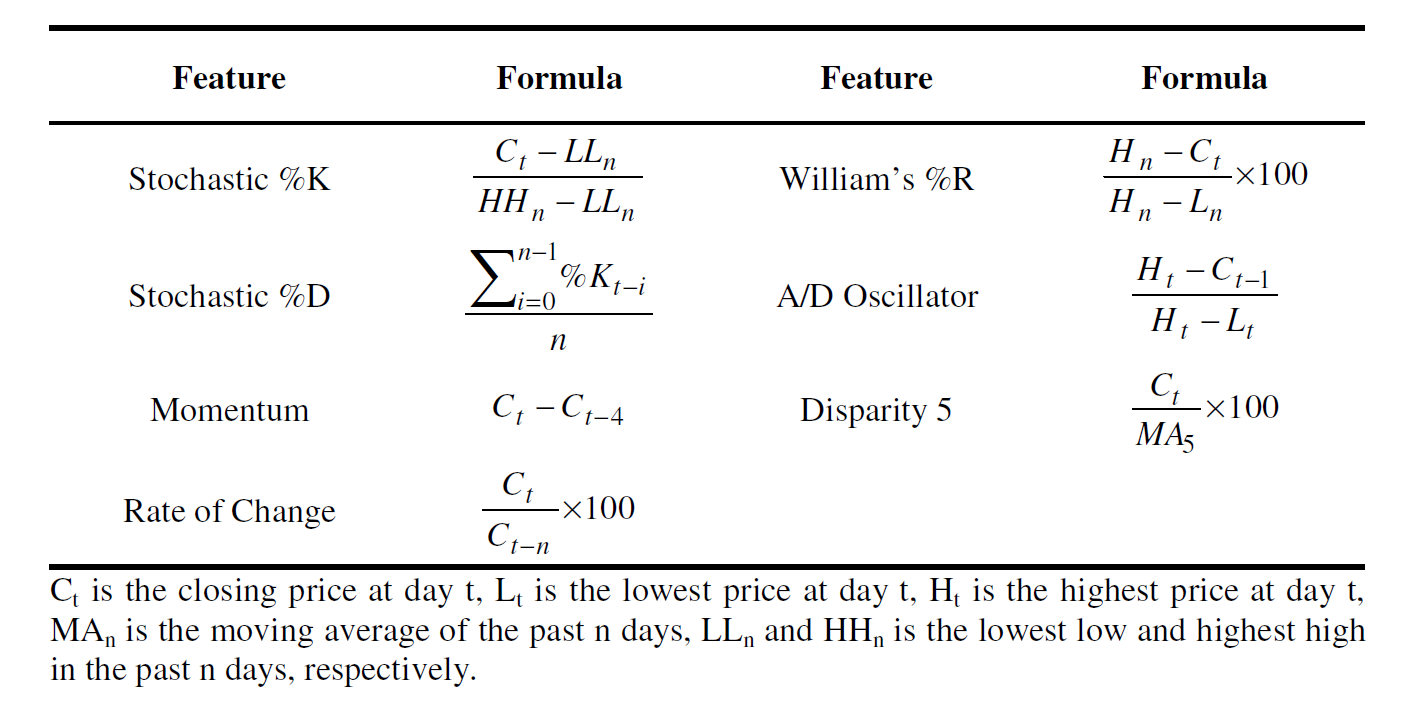}
    \caption{\label{fig:tech}Self attention mechanism}
    \end{figure}
    
    \item \textit{\textbf{CNN-LSTM}}: We use CNN instead of the news level self-attention layer. Note that the word and day level self-attention layer still remains the same. We want to see how well the self-attention layer works compared to CNN which is good at capturing local and semantic information from texts.
    \item \textit{\textbf{E-NN}}: Ding et al.~\cite{ding2014using} reported a system that uses structure event tuples input and standard neural network prediction model.
    \item \textit{\textbf{EB-CNN}}: Ding et al.~\cite{ding2015deep} proposed a model using event embedding input and Convolutional Neural Network as prediction model.
    \item \textit{\textbf{KGEB-CNN}}: Ding et al.~\cite{ding2016knowledge} further incorporated an outside knowledge graph into the learning process for event embedding. The model structure is the same as EB-CNN.
\end{itemize}

\subsection{Result and Discussion}
Shown in table~\ref{tab:Result}, the results on the comparison between the models SVM and the rest of the models indicates that deep neural network model achieves better performance than the SVM model. Comparison between Bag-At-LSTM and At-LSTM demonstrates that sentence encoding with LSTM have slightly better result than Bag-of-words model. Furthermore, WEB-At-LSTM and At-LSTM indicates that the character level composition helps improved the models accuracy. The technical Indicator leveraged from Zhai et al.~\cite{zhai2007combining} doesn't show any performance improvement to our model. In contrast, it results in a decline on the accuracy and it might be caused by adding noise to the dense representation vector $V$ before the output and prediction layer. The comparison between CNN-LSTM and At-LSTM shows that the news level self-attention layer can help capture more relevant news titles and their temporal features. As Ding et al. concluded in~\cite{ding2014using}, the news titles might contain more useful information where as the abstract or article might cause some negative effect to the model. The Ab-At-LSTM and Doc-At-LSTM confirmed this viewpoint since their accuracy are lower than the proposed model that only used the information from the title. 

Our proposed model has a 65.53\% max accuracy and a average accuracy of 63.06\% which is lower than the KGEB-CNN proposed by Ding et al., this is likely due to Knowledge graph event embedding (KGEB) is a more powerful method for model the content in news titles than the sequence embedding shown in this work. 
\begin{table}[H]
\centering
\begin{tabular}{|p{3cm}||p{3cm}|p{3cm}|}
 \hline
 \multicolumn{3}{|c|}{S\&P 500 index prediction Experimental Results} \\
 \hline
 Model & Average Accuracy & Max Accuracy\\
 \hline
SVM & 56.38\% & --\\
\hline
Bag-At-LSTM & 61.93\% & 63.06\%\\
 \hline
WEB-At-LSTM & 62.51\% & 64.42\%\\
 \hline
Ab-At-LSTM & 60.6\% & 61.93\%\\
 \hline
Doc-At-LSTM & 59.96\% & 60.6\%\\
\hline
Tech-At-LSTM & 62.51\% & 64.42\%\\
\hline
CNN-LSTM & 61.36\% & 63.06\%\\
\hline
E-NN & 58.83\% & -- \\
\hline
EB-CNN & 64.21\% & -- \\
\hline
KGEB-CNN & \textbf{66.93}\% & -- \\
\hline
At-LSTM & 63.06\% & \textbf{65.53}\%\\
\hline
\end{tabular}\caption{\label{tab:Result}Experimental Results: We don't know the max accuracy in~\cite{ding2014using,ding2015deep,ding2016knowledge}, so here we assume the accuracy presented in those papers are the average accuracy and use -- to represent the max accuracy.}
\end{table}

We use the At-LSTM model for individual stock price prediction after confirmed that it out performs other approaches. The results are shown in table~\ref{tab:indiResult}, each stock shown in the table has more than 66\% accuracy and the company WALMART has an average accuracy of 70.36\% and max accuracy of 72.06\%. Apparently, predicting individual stock price leads to higher accuracy than predicting S\&P 500 index, and this is mainly due to the news article we used for input. In terms of the individual stock prediction, the news article we used are more relative to it's corresponding company. In contrast, we used the full corpus as input for S\&P 500 index prediction and certainly this adds noise to our model and hence effects the accuracy.
\begin{table}
\centering
\begin{tabular}{|p{3cm}||p{3cm}|p{3cm}|}
 \hline
 \multicolumn{3}{|c|}{Individual stock prediction Experimental Results} \\
 \hline
 Company & Average Accuracy & Max Accuracy\\
 \hline
GOOG & 68.75\% & 71.25\%\\
\hline
AMZN & 67.32\% & 69.46\% \\
\hline
CSCO & 66.82\% & 67.62\%\\
 \hline
MSFT & 67.92\% & 69.89\%\\
\hline
AAPL & 67.52\% & 69.42\%\\
 \hline
INTC & 67.12\% & 67.63\%\\
\hline
IBM & 69.49\% & 71.41\%\\
 \hline
AMD & 66.12\% & 69.10\%\\
\hline
NVDA & 69.35\% & 70.51\%\\
\hline
QCOM & 68.53\% & 69.70\%\\
\hline
WMT& \textbf{70.36}\% & \textbf{72.06}\%\\
\hline
T& 68.53\% & 69.70\%\\
\hline
\end{tabular}\caption{\label{tab:indiResult}Experimental Results for individual stock prediction: We only list the major companies that are in the S\&P 500.}
\end{table}
\section{Conclusion}
This paper has been motivated by the successes of Deep learning methods in Natural Language Processing task. we proposed a Attention-based LSTM model(At-LSTM) to predict the directional movements of Standard \& Poor’s 500 index and individual companies stock price using financial news titles. Experimental results suggests that our model is promising and competitive with the state-of-the-art model which incorporate knowledge graph into the learning process of event embeddings~\cite{ding2016knowledge}.

There are some directions in our future work. While previous work and our result has found that including the body text of the news performs worse than just the headline, there may be useful information to extract from the body text, other directions include looking at predicting price movements at a range of time horizons, in order to gauge empirically how quickly information are absorbed in the market, and relate this to the finance literature on the topic. The financial time series are known by its volatility, in many cases small changes in the series that can be interpreted as noise. Moreover, the elimination of small variations makes the model focus only on news with significant variation on prices which might lead to accuracy increase. 

\bibliographystyle{elsarticle-num}
\bibliography{references}

\begin{thebibliography}{10}
\expandafter\ifx\csname url\endcsname\relax
  \def\url#1{\texttt{#1}}\fi
\expandafter\ifx\csname urlprefix\endcsname\relax\def\urlprefix{URL }\fi
\expandafter\ifx\csname href\endcsname\relax
  \def\href#1#2{#2} \def\path#1{#1}\fi

\bibitem{malkiel1985random}
B.~G. Malkiel, K.~McCue, A random walk down Wall Street, Norton New York, 1985.

\bibitem{wang2012novel}
B.~Wang, H.~Huang, X.~Wang, A novel text mining approach to financial time
  series forecasting, Neurocomputing 83 (2012) 136--145.

\bibitem{basak2007support}
D.~Basak, S.~Pal, D.~C. Patranabis, Support vector regression, Neural
  Information Processing-Letters and Reviews 11~(10) (2007) 203--224.

\bibitem{hearst1998support}
M.~A. Hearst, S.~T. Dumais, E.~Osuna, J.~Platt, B.~Scholkopf, Support vector
  machines, IEEE Intelligent Systems and their applications 13~(4) (1998)
  18--28.

\bibitem{refenes1994stock}
A.~N. Refenes, A.~Zapranis, G.~Francis, Stock performance modeling using neural
  networks: a comparative study with regression models, Neural networks 7~(2)
  (1994) 375--388.

\bibitem{das2012support}
S.~P. Das, S.~Padhy, Support vector machines for prediction of futures prices
  in indian stock market, International Journal of Computer Applications
  41~(3).

\bibitem{lu2009financial}
C.-J. Lu, T.-S. Lee, C.-C. Chiu, Financial time series forecasting using
  independent component analysis and support vector regression, Decision
  Support Systems 47~(2) (2009) 115--125.

\bibitem{malkiel2003efficient}
B.~G. Malkiel, The efficient market hypothesis and its critics, Journal of
  economic perspectives 17~(1) (2003) 59--82.

\bibitem{nofsinger2001impact}
J.~R. Nofsinger, The impact of public information on investors, Journal of
  Banking \& Finance 25~(7) (2001) 1339--1366.

\bibitem{schoneburg1990stock}
E.~Sch{\"o}neburg, Stock price prediction using neural networks: A project
  report, Neurocomputing 2~(1) (1990) 17--27.

\bibitem{akgiray1989conditional}
V.~Akgiray, Conditional heteroscedasticity in time series of stock returns:
  Evidence and forecasts, Journal of business (1989) 55--80.

\bibitem{goccken2016integrating}
M.~G{\"o}{\c{c}}ken, M.~{\"O}z{\c{c}}al{\i}c{\i}, A.~Boru, A.~T.
  Dosdo{\u{g}}ru, Integrating metaheuristics and artificial neural networks for
  improved stock price prediction, Expert Systems with Applications 44 (2016)
  320--331.

\bibitem{adebiyi2014comparison}
A.~A. Adebiyi, A.~O. Adewumi, C.~K. Ayo, Comparison of arima and artificial
  neural networks models for stock price prediction, Journal of Applied
  Mathematics 2014.

\bibitem{kim2012simultaneous}
K.-J. Kim, H.~Ahn, Simultaneous optimization of artificial neural networks for
  financial forecasting, Applied Intelligence 36~(4) (2012) 887--898.

\bibitem{wu2009stock}
D.~Wu, G.~P.~C. Fung, J.~X. Yu, Q.~Pan, Stock prediction: an event-driven
  approach based on bursty keywords, Frontiers of Computer Science in China
  3~(2) (2009) 145--157.

\bibitem{xie2013semantic}
B.~Xie, R.~J. Passonneau, L.~Wu, G.~G. Creamer, Semantic frames to predict
  stock price movement.

\bibitem{ding2014using}
X.~Ding, Y.~Zhang, T.~Liu, J.~Duan, Using structured events to predict stock
  price movement: An empirical investigation, in: Proceedings of the 2014
  Conference on Empirical Methods in Natural Language Processing (EMNLP), 2014,
  pp. 1415--1425.

\bibitem{ding2015deep}
X.~Ding, Y.~Zhang, T.~Liu, J.~Duan, Deep learning for event-driven stock
  prediction., in: Ijcai, 2015, pp. 2327--2333.

\bibitem{ding2016knowledge}
X.~Ding, Y.~Zhang, T.~Liu, J.~Duan, Knowledge-driven event embedding for stock
  prediction, in: Proceedings of COLING 2016, the 26th International Conference
  on Computational Linguistics: Technical Papers, 2016, pp. 2133--2142.

\bibitem{chang2016measuring}
C.-Y. Chang, Y.~Zhang, Z.~Teng, Z.~Bozanic, B.~Ke, Measuring the information
  content of financial news, in: Proceedings of COLING 2016, the 26th
  International Conference on Computational Linguistics: Technical Papers,
  2016, pp. 3216--3225.

\bibitem{peng2015leverage}
Y.~Peng, H.~Jiang, Leverage financial news to predict stock price movements
  using word embeddings and deep neural networks, arXiv preprint
  arXiv:1506.07220.

\bibitem{luss2015predicting}
R.~Luss, A.~d’Aspremont, Predicting abnormal returns from news using text
  classification, Quantitative Finance 15~(6) (2015) 999--1012.

\bibitem{skuza2015sentiment}
M.~Skuza, A.~Romanowski, Sentiment analysis of twitter data within big data
  distributed environment for stock prediction, in: Computer Science and
  Information Systems (FedCSIS), 2015 Federated Conference on, IEEE, 2015, pp.
  1349--1354.

\bibitem{sehgal2007sops}
V.~Sehgal, C.~Song, Sops: stock prediction using web sentiment, in: Data Mining
  Workshops, 2007. ICDM Workshops 2007. Seventh IEEE International Conference
  on, IEEE, 2007, pp. 21--26.

\bibitem{Bag}
Wikipedia, \href{https://en.wikipedia.org/wiki/Bag-of-words_model}{Bag of words
  model} (2014).
\newline\urlprefix\url{https://en.wikipedia.org/wiki/Bag-of-words_model}

\bibitem{Word}
Wikipedia, \href{https://en.wikipedia.org/wiki/Word_embedding}{Word embedding}
  (2015).
\newline\urlprefix\url{https://en.wikipedia.org/wiki/Word_embedding}

\bibitem{mikolov2013efficient}
T.~Mikolov, K.~Chen, G.~Corrado, J.~Dean, Efficient estimation of word
  representations in vector space, arXiv preprint arXiv:1301.3781.

\bibitem{hochreiter1997long}
S.~Hochreiter, J.~Schmidhuber, Long short-term memory, Neural computation 9~(8)
  (1997) 1735--1780.

\bibitem{LSTM}
Wikipedia, \href{https://en.wikipedia.org/wiki/Long_short-term_memory}{Long
  short term memory neural networks} (2015).
\newline\urlprefix\url{https://en.wikipedia.org/wiki/Long_short-term_memory}

\bibitem{sutskever2014sequence}
I.~Sutskever, O.~Vinyals, Q.~V. Le, Sequence to sequence learning with neural
  networks, in: Advances in neural information processing systems, 2014, pp.
  3104--3112.

\bibitem{gers2002learning}
F.~A. Gers, N.~N. Schraudolph, J.~Schmidhuber, Learning precise timing with
  lstm recurrent networks, Journal of machine learning research 3~(Aug) (2002)
  115--143.

\bibitem{kingma2014adam}
D.~P. Kingma, J.~Ba, Adam: A method for stochastic optimization, arXiv preprint
  arXiv:1412.6980.

\bibitem{zeiler2012adadelta}
M.~D. Zeiler, Adadelta: an adaptive learning rate method, arXiv preprint
  arXiv:1212.5701.

\bibitem{Cross}
Wikipedia, \href{https://en.wikipedia.org/wiki/Cross_entropy}{Cross entropy}
  (2014).
\newline\urlprefix\url{https://en.wikipedia.org/wiki/Cross_entropy}

\bibitem{pennington2014glove}
J.~Pennington, R.~Socher, C.~Manning, Glove: Global vectors for word
  representation, in: Proceedings of the 2014 conference on empirical methods
  in natural language processing (EMNLP), 2014, pp. 1532--1543.

\bibitem{chen2017recurrent}
Q.~Chen, X.~Zhu, Z.-H. Ling, S.~Wei, H.~Jiang, D.~Inkpen, Recurrent neural
  network-based sentence encoder with gated attention for natural language
  inference, arXiv preprint arXiv:1708.01353.

\bibitem{kim2014convolutional}
Y.~Kim, Convolutional neural networks for sentence classification, arXiv
  preprint arXiv:1408.5882.

\bibitem{lin2017structured}
Z.~Lin, M.~Feng, C.~N.~d. Santos, M.~Yu, B.~Xiang, B.~Zhou, Y.~Bengio, A
  structured self-attentive sentence embedding, arXiv preprint
  arXiv:1703.03130.

\bibitem{yang2016hierarchical}
Z.~Yang, D.~Yang, C.~Dyer, X.~He, A.~Smola, E.~Hovy, Hierarchical attention
  networks for document classification, in: Proceedings of the 2016 Conference
  of the North American Chapter of the Association for Computational
  Linguistics: Human Language Technologies, 2016, pp. 1480--1489.

\bibitem{zhai2007combining}
Y.~Zhai, A.~Hsu, S.~K. Halgamuge, Combining news and technical indicators in
  daily stock price trends prediction, in: International symposium on neural
  networks, Springer, 2007, pp. 1087--1096.

\end{thebibliography}
\end{document}